\documentclass{aa}

\usepackage{amsmath}
\usepackage{amssymb}
\usepackage{graphicx}
\usepackage{slashed}
\usepackage{color}

\newcommand{\gsim}{\lower.7ex\hbox{$\;\stackrel{\textstyle>}{\sim}\;$}}
\newcommand{\lsim}{\lower.7ex\hbox{$\;\stackrel{\textstyle<}{\sim}\;$}}

\usepackage[normalem]{ulem}

\begin{document}

\title{Simulations of galaxy cluster collisions with a dark plasma component}

\author{
    Christian Spethmann \inst{1}
    \and
    Hardi Veerm\"ae \inst{1}
    \and
    Tiit Sepp \inst{2,3}
    \and
    Matti Heikinheimo \inst{4}  
    \and
    Boris Deshev \inst{2,3} 
    \and
    Andi Hektor \inst{1}
    \and
    Martti Raidal \inst{1}
}

\institute{
    National Institute of Chemical Physics and Biophysics, R\"avala 10, 10143 Tallinn, Estonia
    \and
    Institute of Physics, University of Tartu, W. Ostwaldi 1, Tartu, 50411, Estonia
    \and
    Tartu Observatory, Observatooriumi 1, T\~oravere, Tartumaa, 61602, Estonia
    \and
    Helsinki Institute of Physics, P.O.Box 64, FI-00014 University of Helsinki, Finland
}

\abstract
{Dark plasma is an intriguing form of self-interacting dark matter with an effective fluid-like behavior, which is well motivated by various theoretical particle physics models.}
{We aim to find an explanation for an isolated mass clump in the Abell 520 system, which cannot be explained by traditional models of dark matter, but has been detected in weak lensing observations.}
{We performed $N$-body smoothed particle hydrodynamics simulations of galaxy cluster collisions with a two component model of dark matter, which is assumed to consist of a predominant non-interacting dark matter component and a 10-40\% mass fraction of dark plasma.}
{The mass of a possible dark clump was calculated for each simulation in a parameter scan over the underlying model parameters. In two higher resolution simulations shock-waves and Mach cones were observed to form in the dark plasma halos.}
{By choosing suitable simulation parameters, the observed distributions of dark matter in both the Bullet Cluster (1E 0657-558) and Abell 520 (MS 0451.5+0250) can be qualitatively reproduced.}



\date{\today}


\maketitle

\section{Introduction}

Identifying the nature of dark matter (DM) is the central problem of astroparticle physics. Even though the abundance of DM is about five times as large as that of ordinary baryonic matter, very little is known about the properties of the particles that make up most of our universe. We do not know whether and how DM interacts with itself or with the Standard Model (SM) other than by gravity, or even if DM is made of a single particle species or several different types of particles.

Finding answers to these questions is a crucial first step towards a more complete understanding of Nature. The prevailing models of physics beyond the SM typically predict specific types of DM particles, such as the neutralino or gravitino as the lightest supersymmetric particle. Generally in these models the self-interactions between the DM particles are weak and do not lead to observable effects. The confirmation of DM self-interactions would therefore have a profound impact on the development of theoretical physics not just for our understanding of DM itself, but for physics beyond the SM in general.

Some well-known discrepancies between experimental observations and theoretical predictions support the idea of DM self-interactions, including the cusp-vs-core problem \citep{deBlok:2009sp,Rocha:2012jg} and the missing satellites problem \citep{Klypin:1999uc}. Although it is too early to claim that these problems cannot be solved by baryonic physics, they can potentially be considered as indirect observations of DM self-interactions. On the other hand, cluster mergers offer a possibility to directly observe the effects of DM self-interactions. Whereas observations of the Bullet Cluster (1E 0657-558) \citep{Markevitch2004,Randall:2007ph} and recently several other mergers \citep{Harvey:2015hha} have been used to constrain models of self-interacting DM, the observations of Abell 520 \citep{Mahdavi:2007yp,Jee:2014hja} and Abell 3827 \citep{Massey:2015dkw} seem to suggest that DM indeed exhibits some form of collisional behavior \citep{Heikinheimo:2015kra,Kahlhoefer:2015vua}.

As was argued in \citep{Heikinheimo:2015kra}, multicomponent DM offers an appealing possibility to address the missing satellites and cusp vs.~core problems and to explain the observations of Abell 520 and 3827. Specifically, we assumed that a subdominant component of DM behaves effectively as a collisional fluid, while the main component of DM is collisionless. In \citep{Heikinheimo:2015kra}, the interacting component was taken to consist of fermions charged under a new unbroken $U(1)$ interaction. It therefore forms a dark plasma, reminiscent of the intracluster medium consisting of ionized hydrogen. In this work we aimed to solidify this picture by complementing the previous analysis with detailed $N$-body smoothed particle hydrodynamics (SPH) simulations. We show that the qualitative features observed in the bullet cluster and in Abell 520 can be recovered in the dark plasma model. In Sect.\ref{Dark Plasma} we will review the dark plasma model presented in \citep{Heikinheimo:2015kra}, and justify the use of hydrodynamic $N$-body simulations to model the collisional behavior of the plasma. In Sect.~\ref{Cluster observations} we will outline the features of the observed cluster mergers, and in Sect.~\ref{Simulation} we describe the simulation procedure. We present the results in Sect.~\ref{Results} and our conclusions in Sect.~\ref{Conclusions}.

\section{Dark plasma as a collisional fluid}
\label{Dark Plasma}

Plasma is a state of matter in which collective effects, mediated by long-range interactions, dominate over hard short-range collisions of particles. For the purposes of this work it is not necessary to assume a specific realization of the dark plasma. Our aim is to test a generic scenario, in which dark matter consists of two components. One of the components is assumed to be cold and collisionless, while the second exhibits collisional behavior in a similar way as the intracluster baryonic plasma.  In detail, the simulation therefore addresses any model of dark plasma that has the following properties:
\begin{enumerate}
	\item DM consists of an interacting component and a WIMP-like non-interacting component.
	\item The interacting subcomponent has long-range self-interactions mediated by a dark photon, which render this dark component a plasma unless the coupling is negligibly small.
	\item The radiative cooling of the interacting component is ineffective at astronomical scales.
\end{enumerate}

For concreteness we could consider the minimal model of a dark plasma~\citep{Heikinheimo:2015kra} which consists of a mixture of vector-like dark fermions and antifermions charged under a dark $U(1)$ gauge group. It can be summed up by the Lagrangian
\begin{equation} \label{min_plasma}
	\mathcal{L}
	= \mathcal{L}_{SM} - \frac{1}{4} F_{D\mu\nu}F^{\mu\nu}_{D} + \bar{\chi}\left(i \slashed{D}  - m_{D}\right)\chi,
\end{equation}
where $D_\mu = \partial_\mu - e_{D} A_\mu^{(D)}$ is the covariant derivative, $F_D^{\mu\nu}$ is the field tensor of the dark photon $A^{\mu}_{D}$, $\chi$ is the interacting DM component with mass $m_{D}$ and $e_{D}$ is the dark $U(1)$ charge. For later reference, we also define the dark fine structure constant $\alpha_D = e_{D}^{2}/(4\pi)$.  In the early universe, the dark fermions freeze out via annihilation to the dark photons. If this dark sector was in thermal equilibrium with the visible one at some temperature far above the electroweak scale, then producing the correct relic abundance of dark plasma from freeze-out results in a relation between the mass and charge of the fermions, such that $\alpha_D \sim 10^{-4}$ $(m_D/{\rm GeV})$. However, it is also possible that the the dark plasma is produced by other means, for example by dark freeze-out~\citep{Bernal:2015ova,Chu:2011be}, for which this relation will differ.

Already in the minimal setup it can be seen that the three properties listed above are not very restricting. In the minimal model of Eq.~\eqref{min_plasma} they hold as long as the dark fermion mass is below a TeV, above which the requirement for the correct relic abundance drives the $U(1)$ coupling constant to nonperturbative values. In the following we elaborate on these points in the context of the minimal model for which it is possible to give concrete quantitative results.

The first property that requires DM to contain at least two subcomponents is necessary to meet the observational constraints from the Bullet Cluster that imply that the subcluster has not lost more than 30\% of its mass as it passed through the main halo \citep{Markevitch2004}. It follows that dark plasma can not make up the majority of the subclusters mass which rules out models where all of dark matter is in the state of plasma.

The second property implies that the interacting component of DM cannot freely counter-stream even in the case that the elastic scattering rate is very low and the plasma is seemingly collisionless. This is due to the formation of collisionless shocks \citep{Bret:2015qia} in the event of overlapping bodies of dark plasma with opposite velocities, such as in cluster mergers. These collisionless shocks are generated by the plasma instabilities, and are a prevalent phenomenon in astrophysical plasmas. They have been observed for example in the Earth's bow shock, in the expansion of supernova remnants into the interstellar medium, and in the behavior of the ionized hydrogen in galaxy collisions. The growth rates of these instabilities can be estimated analytically for the dark plasma \citep{Bret:2009fy} and recent numerical results make it possible to estimate the time of shockwave formation. For the minimal model in Eq.~\eqref{min_plasma} the growth rate of the dominant instability mode is of the order of the inverse plasma frequency $\omega_{p}^{-1}$ and a thus conservative order of magnitude estimate of the timescale of shockwave formation \citep{Heikinheimo:2015kra,Bret:2014ufa}, 
\begin{align}
	\tau_{s} 
	&\approx 10^3 \omega_{p}^{-1} \nonumber\\
	&= 7 \, {\rm s} \times  
	\left( \frac{m_{D}}{{\rm GeV} } \right) 
	\left( \frac{\alpha_{D}}{10^{-2}} \right)^{-\frac{1}{2}} 
	\left( \frac{\rho_{\rm DP}}{10^{-2}\, {\rm GeV} / {\rm cm}^{3} } \right)^{-\frac{1}{2}},
\end{align}
is clearly orders of magnitude smaller than the timescales involved in cluster collisions in a wide range of parameters ($\rho_{\rm DP}$ denotes the energy density of the dark plasma). This remains true for most models of charged DM unless the coupling to the dark radiation is negligibly small. Thus, at timescales much longer than the shock formation time, the dark plasma is expected to reach local thermal equilibrium, and therefore behaves effectively as a collisional fluid which can be simulated in a similar manner as the ordinary, visible intergalactic plasma. Indeed, the validity of hydrodynamic simulations in astrophysics relies on the same phenomena taking place in the electron-proton plasma, which in the elastic scattering sense is also collisionless during the relevant timescales~\citep{Bret:2015qia}. At astronomical scales, the dark plasma component can thus be simulated by  hydrodynamical simulation codes.

Scenarios where the dynamics of the interacting particles are dominated by hard short-range collisions have been studied elsewhere \citep{Kahlhoefer:2013dca,Kahlhoefer:2015vua}. In these studies it is typically assumed that DM consists of a single interacting component. Therefore, due to constraints from the Bullet Cluster, it cannot behave as a collisional fluid. The self-scattering rate has therefore to be small enough that the DM does not reach local thermal equilibrium and become an effectively continuous fluid which exhibits pressure. Then, contrary to the case of a plasma, counter-streaming remains possible and the role of the self-interaction is to exert an effective drag force to the bodies of DM in the counter-streaming situation. This effect has been used to explain the separation between the galaxies and the center of mass observed in Abell 3827 \citep{Kahlhoefer:2015vua}, but it cannot generate a separate DM core such as is observed in Abell 520, which seems to require the existence of a multicomponent DM sector, and the possibility for energy dissipation.

In the dark plasma scenario the dynamics of the interacting component are dominated by collective plasma effects instead of the hard self-scatterings processes. In the minimal model \eqref{min_plasma} the mean free path of charged particles due to binary collisions in the plasma is of the order $\lambda_{\mbox{\scriptsize mfp}} \approx 40 \mbox{ kpc} \, \times (m_{D}/\mbox{GeV})$ and this property is therefore clearly satisfied. However, it should be pointed out that for the validity of the simulations performed in this work, it is not strictly required that the effective fluid-like behavior of the interacting component is due to plasma effects. It is conceivable that a multicomponent DM sector contains an interacting component with large enough self-scattering rate, so that this component can reach local thermal equilibrium via hard scattering processes at small scales, and thus forms a collisional gas. We will not pursue the construction of a detailed model of this type here, but simply note that our analysis is valid also in this case.

In the minimal model \eqref{min_plasma}, the characteristic timescale of radiative cooling  is roughly $10^{20}$ yr and therefore negligible. In general, radiative cooling becomes less efficient for larger fermion masses and weaker couplings, since the characteristic timescale of bremsstrahlung cooling is proportional to $m_{D}^{3/2} \alpha_{D}^{-3} $ \citep{Fan:2013yva}. Consequently, the dark plasma cannot cool fast enough to form dark discs or compact astronomical objects such as dark stars. The dissipation of energy via formation of shock waves in dark plasma occurs qualitatively similarly as in baryonic matter \citep{Heikinheimo:2015kra}. In conclusion, dark pair plasma can be simulated as a fluid that does not lose heat through radiative cooling but dissipates energy through shock waves.

The microphysics of our theory is far from unique, since we do not fix the specific model of the dark plasma. Even in the minimal model presented in \eqref{min_plasma}, a wide range of masses and coupling constants are viable. The physical coefficient of shear viscosity of a fully ionized, unmagnetized plasma is given by \citep{Sijacki:2006zua}
\begin{equation} \label{phys_visc}
	\eta_{\mathrm{ph}} = 0.406 \; \frac{ m_{D}^{1/2} T^{5/2}}{ \alpha_D^2 \, \log \Lambda},
\end{equation}
where $\Lambda$ is the number of particles within a Debye sphere and $T$ the temperature. The coefficient of bulk viscosity of such a plasma is zero. We note that natural units $\hbar = c = k_B = 1$ are used here.

The artificial viscosity parameter in SPH simulations does not directly correspond to the physical viscosity of the plasma \citep{Sijacki:2006zua}. Instead, it is introduced because SPH codes lack the ability to sharply resolve hydrodynamic instabilities \footnote{This concept is related to the eddy viscosity in the context of modeling turbulent behavior of fluids.}. The artificial viscosity thus models the shocking behavior by turning kinetic energy of counter-streaming plasmas into heat. We therefore do not fix the value of the artificial viscosity in our simulations and scan over a range of physical values. In this way it is possible to find the value of the artificial viscosity which best corresponds to observations. 
  
\section{Abell 520 and the Bullet Cluster}
\label{Cluster observations}
  
So far only few cluster collisions that contain shock fronts with Mach numbers significantly larger than one are known~\citep{Markevitch2006, Botteon:2017gdi}. These rare objects can provide invaluable insights into the dynamics of dark matter. The better-known cluster of this type is the Bullet cluster (1E 0657-56) at redshift $z=0.29$. The other, Abell 520 (MS 0451.5+0250), is a major cluster merger at redshift $z=0.201$ observed after core passage \citep{Mann:2011kh}. These two clusters exhibit a clear separation between the hot intra-cluster gas and the galaxies, and in the case of Abell 520 a dark core nearly devoid of galaxies~\citep{Mahdavi:2007yp,Jee:2012,Jee:2014hja}. Some other examples of high Mach number cluster collusions, for example A2146 or MACS J0417.5-1154~\citep{Mann:2012aaa}, have larger uncertainties and are left for future studies.

The merger of the two most massive clumps in A520 progresses approximately in the plane of the sky in a NE-SW direction. This picture is supported by the presence of a shock front in the X-ray emitting gas residing between the two clumps \citep{Markevitch2005}. From the geometry of the Mach cone \citep{Markevitch2005} in A520 it is possible to estimate the shock velocity as 2300 km/s (Mach number  $2.2^{+0.9}_{-0.5}$ ) and the time since core passage as $\sim$ 1 Gyr. It was estimated that the velocity of the impact in the Bullet cluster is $4700\pm630$ km/s (Mach number $3.0\pm0.4$) while the time since core passage is approximately 0.15 Gyr~\citep{Markevitch2006}.

A series of multi-object spectroscopy observations in the field of A520 indicate a total mass of 11.6 $^{+3.7}_{-3.0} \times 10^{14} M_{\odot}$ and a velocity dispersion of 1036$^{+101}_{-97}$ estimated from the peculiar velocities of 315 cluster members \citep{2017arXiv170703208D}. This estimate is broadly consistent with the X-ray temperature \citep{Govoni2004}; the total mass of the cluster was estimated at $9.5\times10^{14} \; M_{\odot}$ \citep{ChungSM2009}.  A series of weak lensing analyses of A520 \citep{Mahdavi:2007yp,Okabe:2007af,Jee:2012,Jee:2014hja,Clowe2012} reveals a complex structure with two major mass concentrations, both with individual masses around $4 \times 10^{13} \; h^{-1}_{70} \; M_{\odot}$, and at least two other, smaller mass concentrations residing on an axis approximately perpendicular to the main axis in the plane of the sky. All but one \citep{Clowe2012} of the above mentioned analyses confirm the existence of a dark core ($M/L \geq 800 \; M_{\odot}/L_{\odot}$) that is positioned between the two main clumps and coincides with the peak of the X-ray emission, where the stellar light concentration is low \citep{Mahdavi:2007yp,Jee:2012,Jee:2014hja}.  The fact that the position of the dark core coincides with the X-ray emitting gas is the main motivation of our hypothesis that a sub-component of dark matter might be dynamically similar to the baryonic plasma.

According to \citep{Jee:2014hja} the mass within an aperture with radius 150 kpc centered on the dark peak is $M_{\mathrm{ap}}(r<150 \mbox{ kpc}) = 3.94 \pm 0.30 \times 10^{13} \; h^{-1}_{70} \; M_{\odot}$. The DM self-interaction cross section is estimated as $\sim 3.8$ cm$^2$ g$^{-1}$, based on the mass of the separate components in A520 \citep{Mahdavi:2007yp}. For comparison, an upper limit for the same parameter at $<$1 cm$^2$ g$^{-1}$ has been found based on observations of the Bullet cluster \citep{Markevitch2004}. A refined weak-lensing analysis based on Hubble Space Telescope imaging confirms the presence of a central dark clump in A520, but states that the self-interaction cross-section can be brought into agreement with the Bullet cluster value if a longer path is allowed through the separate substructures \citep{Jee:2014hja}. The weak lensing analysis of the Bullet cluster reveals a somewhat simpler picture with only two clumps merging approximately in the plane of the sky \citep{Bradac2006}.

For completeness we remark that a high velocity group separated from the main system by 2000 km s$^{-1}$ was detected in A520 \citep{Girardi:2008}. This discovery led to the speculation that the merger might contain yet another axis, aligned with the line of sight, thus providing an alternative explanation for the dark mass concentration. In this article we will not consider this scenario and instead focus on dark matter self-interactions as a possible explanation.

\section{Simulation and initial conditions}
\label{Simulation}

The aim of the simulation is to provide a proof of concept that the central clump in the Abell 520 cluster can be explained if a sub-component of dark matter behaves like a collisional fluid on macroscopic scales. To test this dark plasma model we performed a series of simulations over Abell 520-like clusters with idealized initial conditions. The simulations were performed using {\sc Gadget-2} \citep{Springel:2005}, an $N$-body/SPH code that has been applied for a wide variety of astrophysical simulations  \citep{Regan:2006fs,Dolag:2008ya,Hou:2012xq,McCracken:2014cka}, in particular to simulate various cluster mergers \citep{Springel:2007,Machado:2013jq, Lage:2014}. The effects of shock mechanics and hydrodynamics are similarly represented in different cosmological simulation codes, which therefore yield equivalent results \citep{Vazza:2011, Sijacki:2012}.

The non-interacting DM component was simulated using the $N$-body dynamics of {\sc Gadget-2}, while the dark plasma was treated as a standard {\sc Gadget-2} SPH component. To test the hypothesis put forward in this paper we considered different values of the artificial viscosity of the dark plasma component and varied its mass fraction. This idealised two component approach provides a clean environment to understand the dark plasma effects. 

Simulations using artificial viscosity can be carried out by alternative SPH codes such as {\sc Vine} \citep{Wetzstein:2009} or {\sc Gizmo} \citep{Hopkins:2014}  in a straightfoward way. For adaptive mesh refinement (AMR) codes, which use an Eulerian grid instead of following the fluid elements in a Lagrangian manner, dark plasma simulations need to be carried out differently, since they generally do not contain any artificial viscosity term (the known exception being {\sc Enzo/Zeus} \citep{Bryan:2014}).  AMR simulations using codes such as {\sc Enzo} \citep{Bryan:2014} or {\sc Ramses} \citep{Teyssier:2002} should instead be conducted with the physical value of the shear viscosity as predicted by the microphysical model, which depend on the local density and temperature of the plasma as stated in Eq.~\eqref{phys_visc}.

As initial conditions for our simulations we constructed spherically symmetric initial halos following Navarro{\textendash}Frenk{\textendash}White (NFW) profiles \citep{NFW:1996}. The density profile of our halos is 
\begin{equation}
	\rho(r) = \left\{ 
            \begin{array}{ll} 
                \displaystyle \frac{\rho_0}{r/r_S \; (1+ r/r_S)^2} & \qquad \mbox{if $r<3 \, r_{200}$}  \rule[-3ex]{0ex}{2ex}\\
                \displaystyle 0 & \qquad \mbox{if $r>3 \, r_{200}$} 
       \end{array} \right.  
\end{equation}
where $\rho(r)$ is the radial DM density function, $\rho_0$ is a multiplicative constant and $r_S$ is the scale radius of the halo. The virial radius $r_{200}$ is defined as the radius where the local density is 200 times above the critical density of the universe, which we take as $\rho_{\rm crit} = 10^{-26} \mbox{ kg/m$^3$} \approx 147 \; M_{\odot} / {\rm kpc}^3$. The scale radius $r_S$ is related to the virial radius by $r_{200} = c \, r_S$, where $c$ is the concentration parameter of the halo. The virial mass of the halo reads
\begin{equation}
	M_{200} = 4 \pi \rho_0 \, {r_S}^3 \left( \log(1+c) - \frac{c}{1+c} \right) .
\end{equation}
We extend our halos to a cutoff radius of $3 r_{200}$ in order to reduce the discontinuity at the edge of the halos. The total halo mass is therefore larger than the virial mass by
\begin{equation}
M_{\rm tot} = \frac{\log(1+3c)-3c/(1+3c)}{\log(1+c)-c/(1+c)} \; M_{200}  .
\end{equation}
and the gravitational potential of the halo reads
\begin{align}
 & \Phi(r) = \\ \nonumber
 & -G \, \frac{M_{\rm tot}}{3 r_{200}} 
 + 4 \pi \rho_0 r_S^2 G 
 \left( \frac{\log (1+3c)}{3c} - \frac{\log (1+r/r_S)}{r/r_S} \right) 
\end{align}
within the halo (that is for $r<3 \, r_{200}$) and
\begin{equation}
\Phi(r) = - G \, \frac{M_{\rm tot}}{r} 
\end{equation}
outside the halo (that is for $r>3 \, r_{200}$).

The initial velocity distribution of the DM particles at a given distance is chosen to follow a Tsallis distribution~\citep{Lisanti:2010qx}
\begin{equation}
	f(v) \propto (1- v^{2}/v_{\rm esc}^{2})^{\alpha}
\end{equation}
where $v_{\rm esc} = \sqrt{-\xi \Phi}$ denotes the local escape velocity and $\alpha = 2$. We reduce the kinetic energy to at most $\xi = 90\%$ of the negative potential energy; this reduction by 10\% practically eliminates the possibility that particles escape the cluster and therefore leads to a more stable initial state. The directions of the initial velocities are distributed uniformly.

The plasma subhalos are constructed by setting initial velocities to zero and imposing hydrostatic equilibrium, while the density follows the same NFW halo profile as the noninteracting dark matter. To this end, the temperature profile is chosen such that the pressure at any radius exactly balances the gravitational forces from both the plasma and the ordinary dark matter components.

The viability of these initial conditions was tested by simulating isolated haloes for a period of 8 Gyr.  The velocity distributions (including the kinetic energy reduction factor $\xi$) were chosen such that after a relatively brief relaxation time of $\sim$1 Gyr the particles in the simulation reach a stable phase space distribution. The initial conditions for our simulations were then chosen such that the halos have sufficient time to relax to their equilibrium states before the collision event.

To find the correct concentration parameters and masses for our halos, we performed a set of simulations (with 8$\cdot 10^4$ particles in each) and compared their final states with observational data. In Ref.~\citep{Jee:2014hja} several compact subhalos of Abell 520 are listed. The projected masses of these subhalos within a radius of 150 kpc were found to be in the order of $5 \cdot 10^{13} \; M_\odot$. The two main subhalos are separated by a distance  $d_{\rm final} \approx$ 1000 kpc. To find the correct initial conditions for our simulation we constructed intial NFW halos with a range of different masses and concentration parameters. By running simulations for each set of parameters we found that an initial halo mass of $4 \cdot 10^{14} \; M_\odot$ and a concentration parameter of 4.0 yields the final state that most closely resembles the observations. The detailed results of our initial condition parameter scan are listed in Table \ref{tab:init_scan}.

\begin{figure*}
   \centering
   \includegraphics[width=0.6\textwidth]{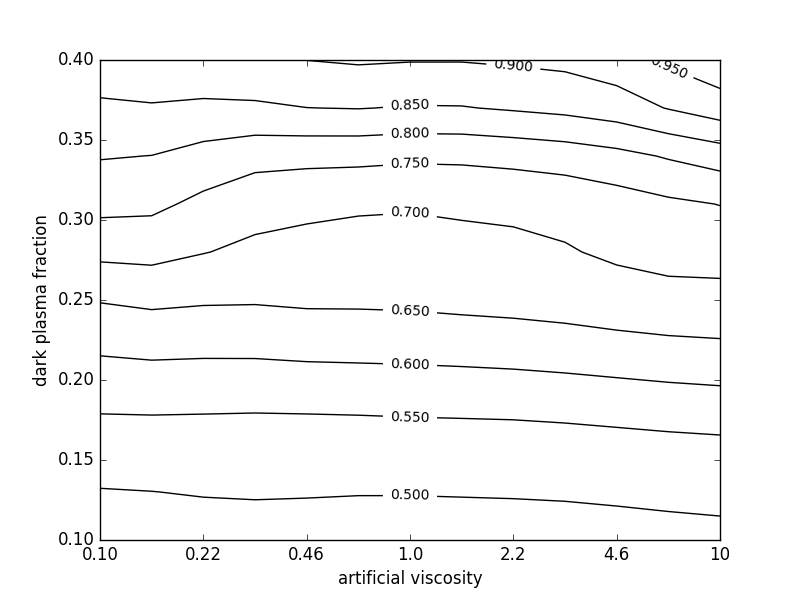}
   \caption{Contour plot of the mass of the central subhalo divided by the average mass of the outer subhalos, $R_C$ . All simulations were done with an impact parameter of $b=150$ kpc.}
   \label{fig:cluster_ratio}
\end{figure*}

\begin{table}
\caption{\label{tab:init_scan} Resulting subhalo parameters from our scan over the initial masses and concentration parameters in the Abell 520 cluster. For illustration we also list the resulting virial radii of the initial clusters. We find that colliding two NFW subhalos with initial masses of $4.0 \cdot 10^{14} M_\odot$ and concentration parameters of 4.0 yields the final state which most closely resembles observations.}

\small
\begin{tabular}{ccccc} 
$M_{\rm init} / M_\odot$ \rule{1ex}{0ex} & $c_{\rm init}$ &  \rule{1ex}{0ex} $R_{200}$ \rule{1ex}{0ex} & \rule{1ex}{0ex} $M_{150}/M_\odot$ \rule{1ex}{0ex} & \rule{1ex}{0ex} $d_{\rm final}$ \rule{1ex}{0ex} \rule{0ex}{4ex} \\ \hline \hline
$3.0 \cdot 10^{14}$ & 4.0 & 861 kpc & $4.96 \cdot 10^{13}$ & \rule{0ex}{0ex} 700.3 kpc \rule{0ex}{3ex} \\
$3.0 \cdot 10^{14}$ & 4.5 & 848 kpc & $5.28 \cdot 10^{13}$ & 762.4 kpc \\
$3.0 \cdot 10^{14}$ & 5.0 & 836 kpc & $5.62 \cdot 10^{13}$ & 720.0 kpc \\ \hline
$3.5 \cdot 10^{14}$ & 4.0 & 906 kpc & $5.05 \cdot 10^{13}$ & \rule{0ex}{0ex} 900.2 kpc \rule{0ex}{3ex} \\
$3.5 \cdot 10^{14}$ & 4.5 & 893 kpc & $5.36 \cdot 10^{13}$ & 860.2 kpc \\
$3.5 \cdot 10^{14}$ & 5.0 & 881 kpc & $5.29 \cdot 10^{13}$ & 782.3 kpc \\ \hline
$4.0 \cdot 10^{14}$ & 4.0 & 948 kpc & $4.82 \cdot 10^{13}$ & \rule{0ex}{0ex} 961.9 kpc \rule{0ex}{3ex} \\
$4.0 \cdot 10^{14}$ & 4.5 & 936 kpc & $5.41 \cdot 10^{13}$ & 821.0 kpc \\
$4.0 \cdot 10^{14}$ & 5.0 & 921 kpc & $5.10 \cdot 10^{13}$ & 900.2 kpc \\ \hline
$4.5 \cdot 10^{14}$ & 4.0 & 986 kpc & $4.94 \cdot 10^{13}$ & \rule{0ex}{0ex} 941.9 kpc \rule{0ex}{3ex} \\
$4.5 \cdot 10^{14}$ & 4.5 & 971 kpc & $5.20 \cdot 10^{13}$ & 860.2 kpc \\
$4.5 \cdot 10^{14}$ & 5.0 & 958 kpc & $5.49 \cdot 10^{13}$ & 821.0 kpc \\
\end{tabular}

\end{table}

\begin{figure*}
   \centering
   \includegraphics[width=0.49\textwidth]{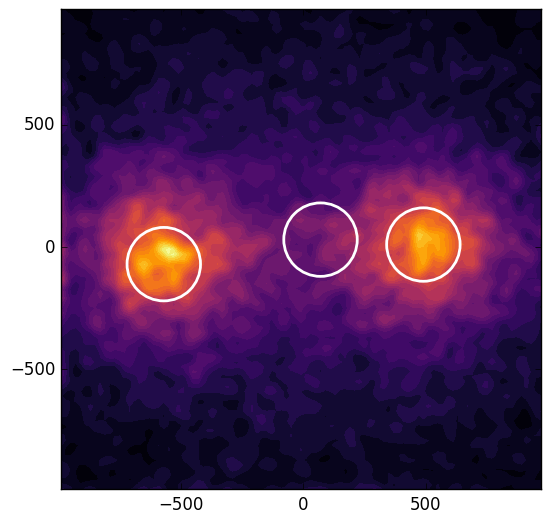}
   \includegraphics[width=0.49\textwidth]{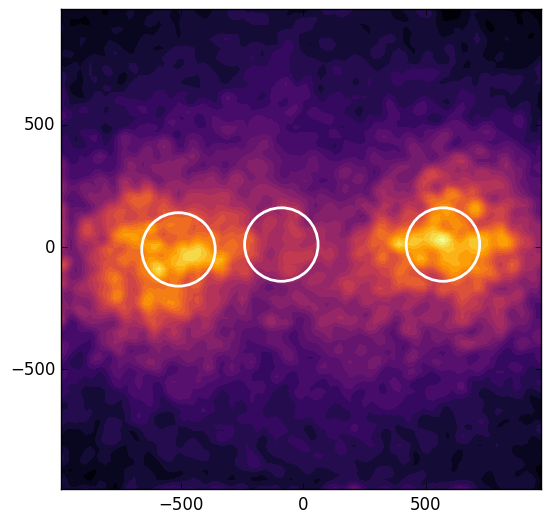}
   \includegraphics[width=0.49\textwidth]{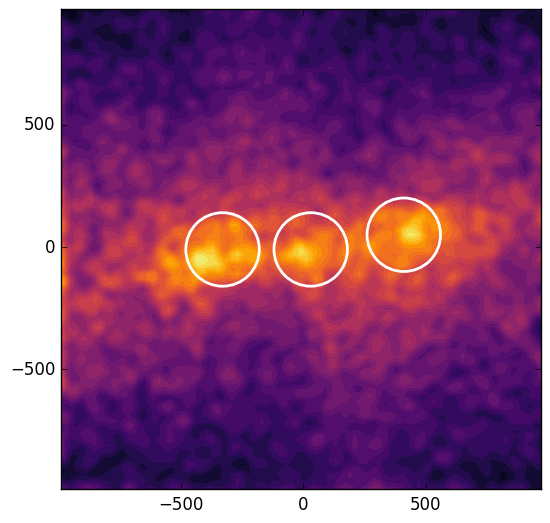}
   \includegraphics[width=0.49\textwidth]{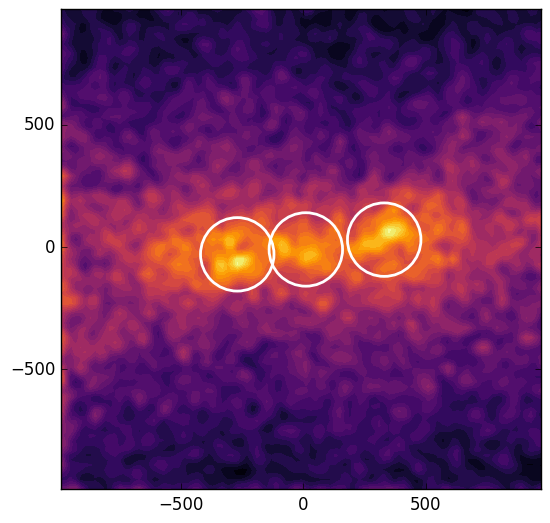}
   \caption{Results of the halo-finding algorithm for four different simulation runs. Top left: $f_{\rm DP}=10\%$ and $\eta=1.0$. In this case there is no central subhalo. Top right: $f_{\rm DP}=25\%$ and $\eta=1.0$. The algorithm found a small subhalo resulting from the Mach cones of the interacting dark plasma component. Bottom left: $f_{\rm DP}=40\%$ and $\eta=1.0$. In this case, there is a clear central subhalo. Bottom right: $f_{\rm DP}=40\%$ and $\eta=10.0$. In this case a central subhalo is present, and the distance of the outer subhalos is reduced. The scale of all pictures is given in kpc, and the contours represents the projected mass density.}
   \label{fig:Clusters}
\end{figure*}

The simulation of the Bullet Cluster was performed with two isolated halos that match the ones used in~\citep{Springel:2007}:  the masses are $M_{200} = 1.5 \times 10^{15} \, M_\odot$ and $M_{200} = 1.5 \times 10^{14} \, M_\odot$ and the concentration parameters for the two initial halos were $c = 2.0$ and $c = 7.0$, the virial radii $r_{200} = 1600$ kpc and $r_{200} = 635$ kpc. We chose a dark plasma mass fraction $f_{\mathrm{DP}} = 0.25$ and an impact parameter of $b = 150$ kpc. The initial relative velocity was taken to be 3500 km/s and the initial separation of the halos was 10 Mpc. This results in a relative velocity of 5200 km/s at the time of core passage which is agreement with the estimate $4700\pm630$ km/s.

\begin{figure*}
   \centering
   \includegraphics[width=\textwidth]{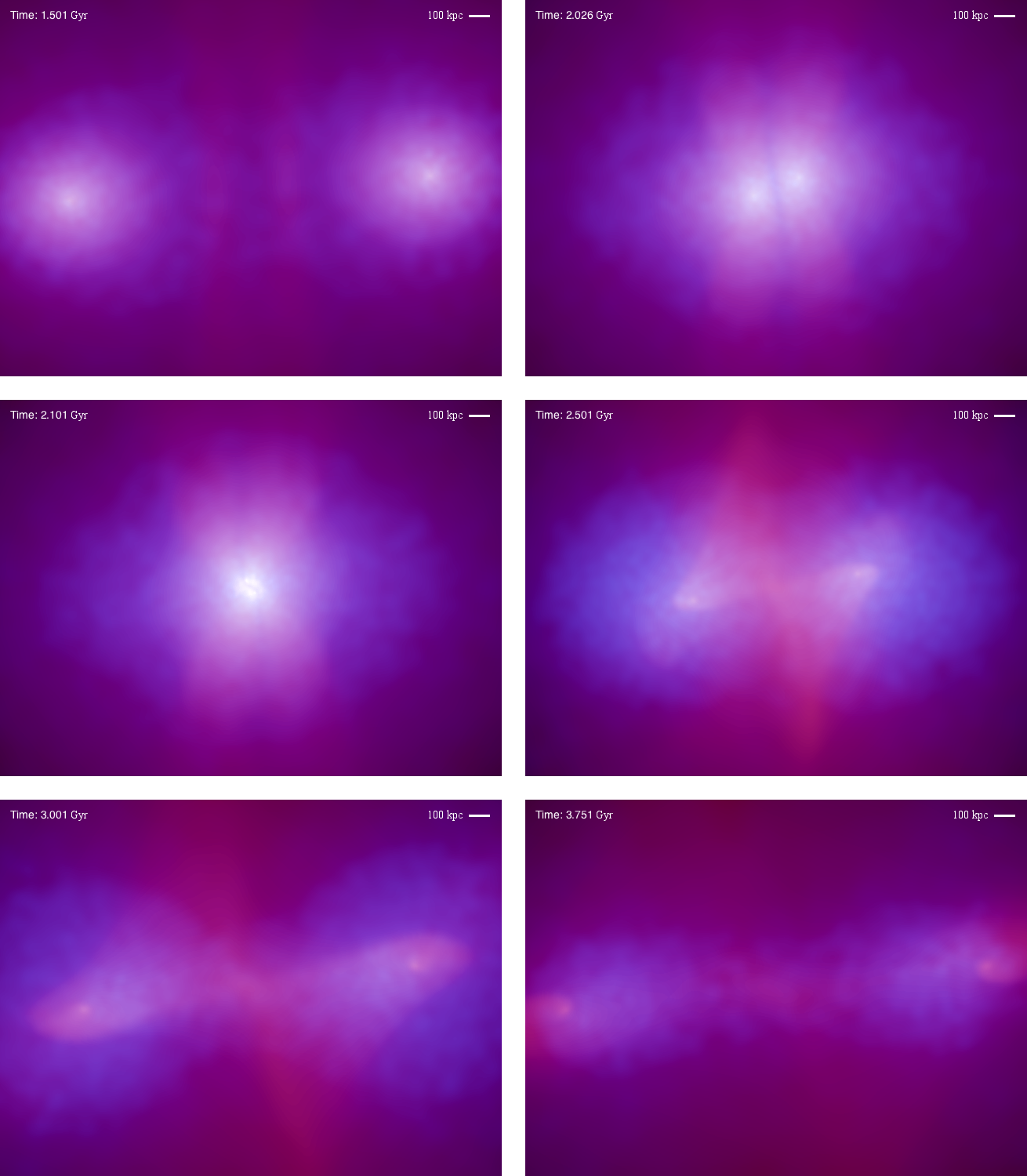} 
   \caption{Simulation of Abell520 with $f_{\rm DP}=25 \%$ and $\eta = 0.7$. The non-interacting DM distribution is displayed in blue and the dark plasma distribution in red. The  color intensities represent the projected mass densities of the two subcomponents. The field of view is 2.35 Mpc $\times$ 1.76 Mpc. The corresponding video is available at \protect\url{http://coe.kbfi.ee/pmwiki/pmwiki.php/Results/Results} }
   \label{fig:videotrain}
\end{figure*}

\begin{figure*}
   \centering
   \includegraphics[width=\textwidth]{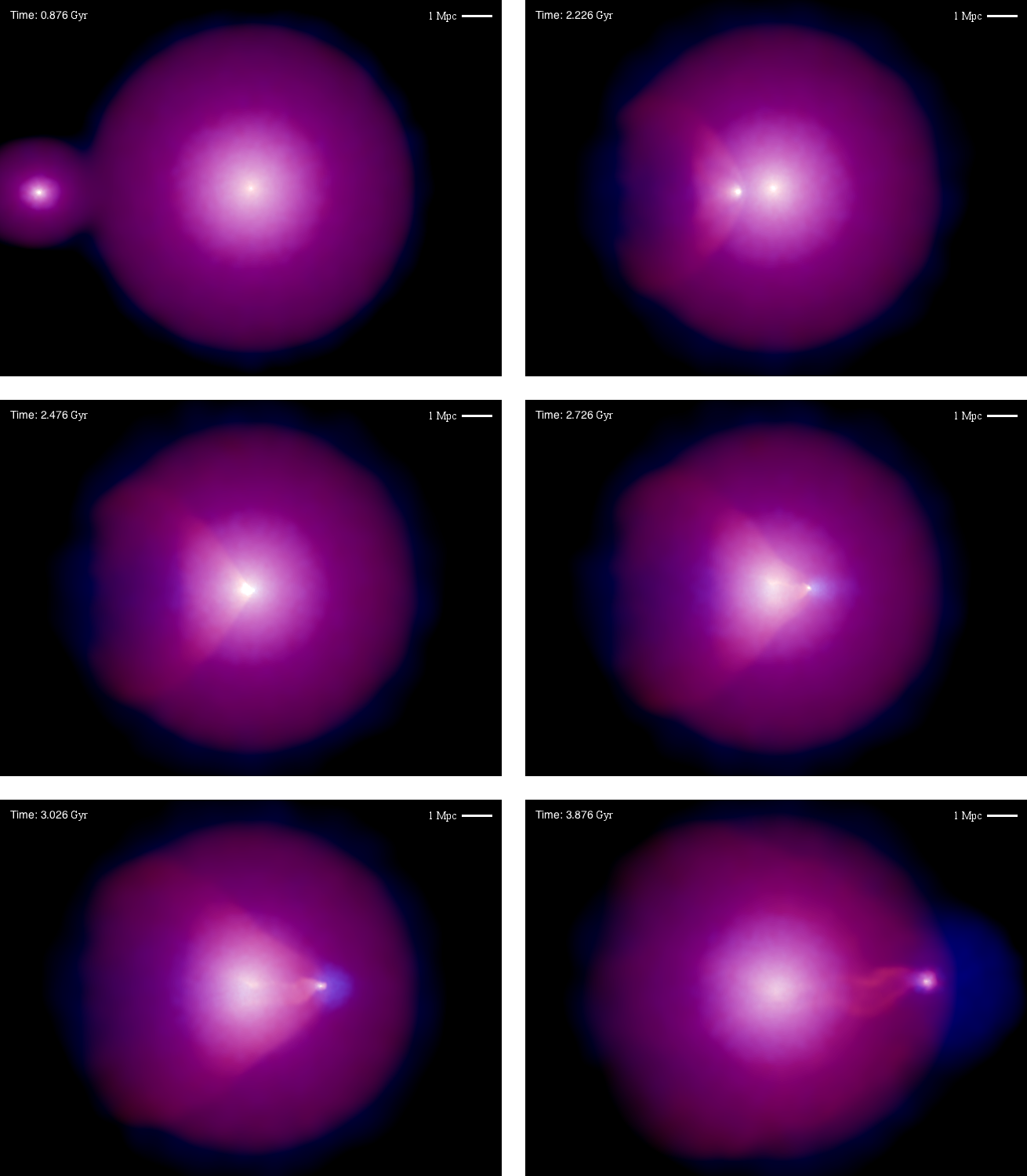}
   \caption{Simulation of the Bullet Cluster with $f_{\rm DP}=25 \%$ and $\eta = 0.7$. The non-interacting DM distribution is displayed in blue and the dark plasma distribution in red. The color intensities represent the projected mass densities of the two subcomponents. The field of view is 19.4 Mpc $\times$ 14.5 Mpc. The corresponding video is available at \protect\url{http://coe.kbfi.ee/pmwiki/pmwiki.php/Results/Results} }
   \label{fig:videobullet}
\end{figure*}

\section{Results}
\label{Results}

To obtain the main results of our paper we carried out a scan in which we kept the concentration parameters and the halo masses fixed. The initial impact parameter was set at 150 kpc, the initial distance between the cluster centers at 6 Mpc (such that the initial halos did not overlap) and the initial relative velocity as 2300 km/s. The dark plasma mass fraction was varied from $f_{\rm DP} = 0.1$ to 0.4 and the artificial viscosity parameter of the dark plasma was varied on a logarithmic scale between $\eta = 0.1$ to 10. A total of 143 simulations were performed. In each simulation run of the scan, 8$\cdot 10^4$ particles were included, with one half representing the dark plasma and the other half the non-interacting DM component.  

In addition, we performed a control simulation with a higher resolution of $8\cdot 10^5$ particles with $f_{\rm DP} = 0.25$ and $\eta = 0.7$. It was found that the lower resolution scan yielded identical results and is therefore sufficient to reproduce the relevant features of the final state.

To check for the presence of a central mass clump in the final state of our simulations and to quantify the result, it was necessary to develop an algorithm that automatically analyses {\sc Gadget2} snapshot files. The algorithm first calculates a projection of the mass distribution onto the plane of the sky and then searches for three local maxima of the projected density distribution in the left, central and right area of the observation plane. Here a maximum is defined as a region in which a circle of 150 kpc radius contains the largest projected mass. 

In Fig.~\ref{fig:Clusters} we show the results of the halo-finding algorithm in four individual simulations of our parameter scan. To quantify the results of our scan, we identify the {\sc Gadget2} snapshot file in which the distance between the two outer subhalos is closest to 1 Mpc and calculate the fraction of the subhalo masses
\begin{equation} 
R_C = \frac{2 M_C}{M_1 + M_2}, 
\end{equation}
where $M_C$ is the mass of the central subhalo and $M_1, M_2$ are the masses of the two outer subhalos. $R_C$ therefore corresponds to the mass of the central subhalo divided by the average mass of the two outer subhalos. 

In Fig.~\ref{fig:cluster_ratio} we present the value of this ratio as a function of the dark plasma fraction $f_{\rm DP}$ and the artificial viscosity parameter $\eta$. We find that the mass fraction $R_C$ is mostly dependent on the plasma fraction and is relatively insensitive to the artificial viscosity as we vary it over 2 orders of magnitude. However, its dependence on the viscosity increases with increasing plasma fraction.

For small values of the dark plasma fraction, we find that the mass of the inner subhalo is approximately 50\% of the outer subhalo mass and can thus be explained by the overlap of the outer halos. In this case, the algorithm finds the projected mass resulting from the overlap of the two DM halos, even if there is actually no subhalo present in the center. For intermediate values of the dark plasma fraction, around 25\%, the central mass is increased by the presence of a concentration of interacting fluid, which forms Mach cones as seen in the full resolution simulation  depicted below in Fig.~\ref{fig:videotrain}. For dark plasma fraction values near the maximum of our simulation of $f_{\rm DP} = 40 \%$, the mass of the central subhalo approaches the mass of the outer subhalos. We find that the final state of the collision becomes more compact for larger values of the artificial viscosity parameter, such that the final separation of the outer cores does not increase beyond 800 kpc. In this case, the mass of the central subhalo increases even more due to the larger overlap between the outer dark plasma subhalos and the center.

Increasing the viscosity results in a larger concentration of dark plasma in the center of the halo, which in turn has an impact on the velocities of the outer clusters. As the dark plasma is slowed down it exerts a decelerating force on these clusters. This effect can be observed in Fig.~\ref{fig:Clusters}. The bottom right panel depicts a simulation with a large viscosity, $\eta = 10$ and $f_{\rm DP} = 0.4$. In this panel the outer clusters have reached their maximal distance of 0.84 Mpc and begin the infall towards the center. This situation should be contrasted with the top left panel that has a smaller plasma fraction  $f_{\rm DP} = 0.1$ and a smaller viscosity  $\eta = 1$. The outer halos are separated by 1Mpc and continue moving apart. It would be interesting to study this effect in the Abell 3827 cluster.

In Fig.~\ref{fig:videotrain} we show individual frames of our high-resolution control simulation with $f_{\rm DP} = 0.25$ and $\eta=0.7$. Panel 1 shows the DM distribution 600 million years before core passage. The two spherical halos move towards each other with a velocity of 2300 km/s and an impact parameter of 150 kpc. Panel 2 depicts the distribution 75 Myr before core passage. The dark plasma component has formed an initial shock front in the space between the cores. Panel 3 shows the moment of core passage. The transversely aligned dark plasma shock is clearly visible. In panel 4 the state 400 Myr after core passage is portrayed. The dark plasma has formed Mach cones, while the non-interacting components follow gravitationally deflected trajectories. In panel 5, 900 Myr after core passage, the two dark matter halos have become more diffuse and the motion away from the center has almost stopped. The dark plasma component is spread between the two cores. Panel 6 shows the state of the collision 1650 Myr after core passage. The two cores have stopped their motion away from each other and begin to fall back to the center. 

As a consistency check we simulated the Bullet Cluster, assuming an identical composition of DM as for the simulation of Abell 520 depicted in Fig.~\ref{fig:videotrain}, i.e.~viscosity $\eta = 0.7$ and dark plasma fraction $f_{\rm DP}=25 \%$. The time evolution of the Bullet Cluster is depicted in Fig.~\ref{fig:videobullet}. Panels 1 and 2 represent snapshots before the core impact. In panel 2 the dark plasma forms a pronounced bow shock. Panel 3 represents the moment of core passage. As the collision progresses, a significant fraction of the dark plasma component of the smaller cluster is absorbed by the larger halo, as shown in panels 4-6, depicting times after the collision. Panel 5 corresponds to the current observation of the Bullet Cluster. After the collision the dark plasma fraction in the bullet is reduced to 17\%. The results of the simulation are thus consistent with all existing observational constraints \citep{Markevitch2004}.

The cone-shaped structures during core passage are reminiscent of cold fronts, which have been observed in the visible intracluster medium in similar collision events~\citep{ZuHone:2016put}. Such cold fronts are expected to form if the core of the initial gas profile is at a lower temperature than the outer regions of the gas halo. The exact nature of the shock fronts thus depends on the unknown initial halo profile of the dark plasma. 

\begin{figure*}
   \centering
   \includegraphics[width=\textwidth]{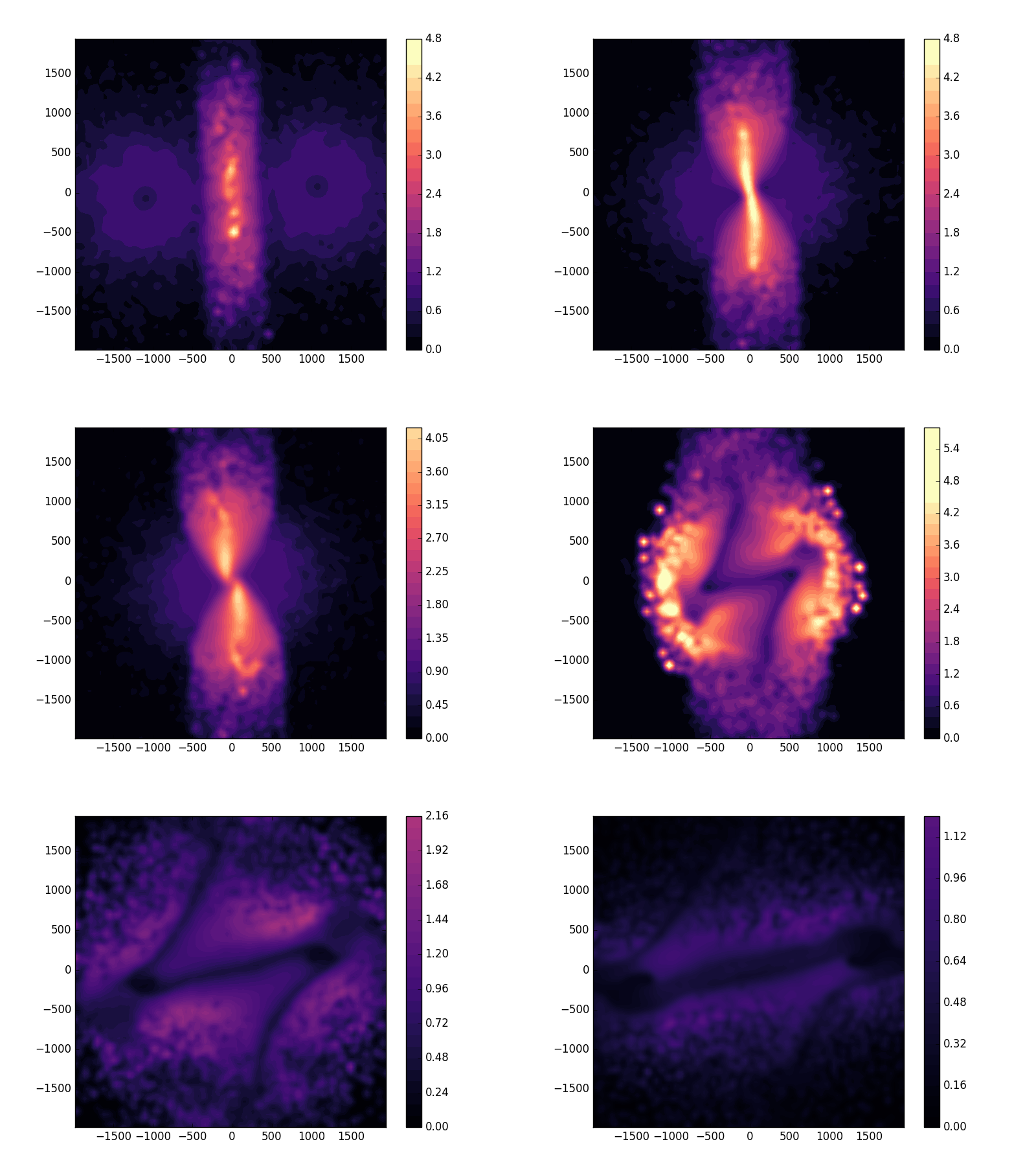} 
   \caption{Internal energy of the dark plasma in a central slice of the Abell 520 collision. The epochs of the panels correspond to the same epochs in Fig.~\ref{fig:videotrain}. All distances are in kpc, the internal energy per unit mass is given in units of 10$^6$ (km/s)$^2$. The energy scale is chosen such that the same colours correspond to equal temperatures in all panels.}
   \label{fig:abell_T}
\end{figure*}

\begin{figure*}
   \centering
   \includegraphics[width=\textwidth]{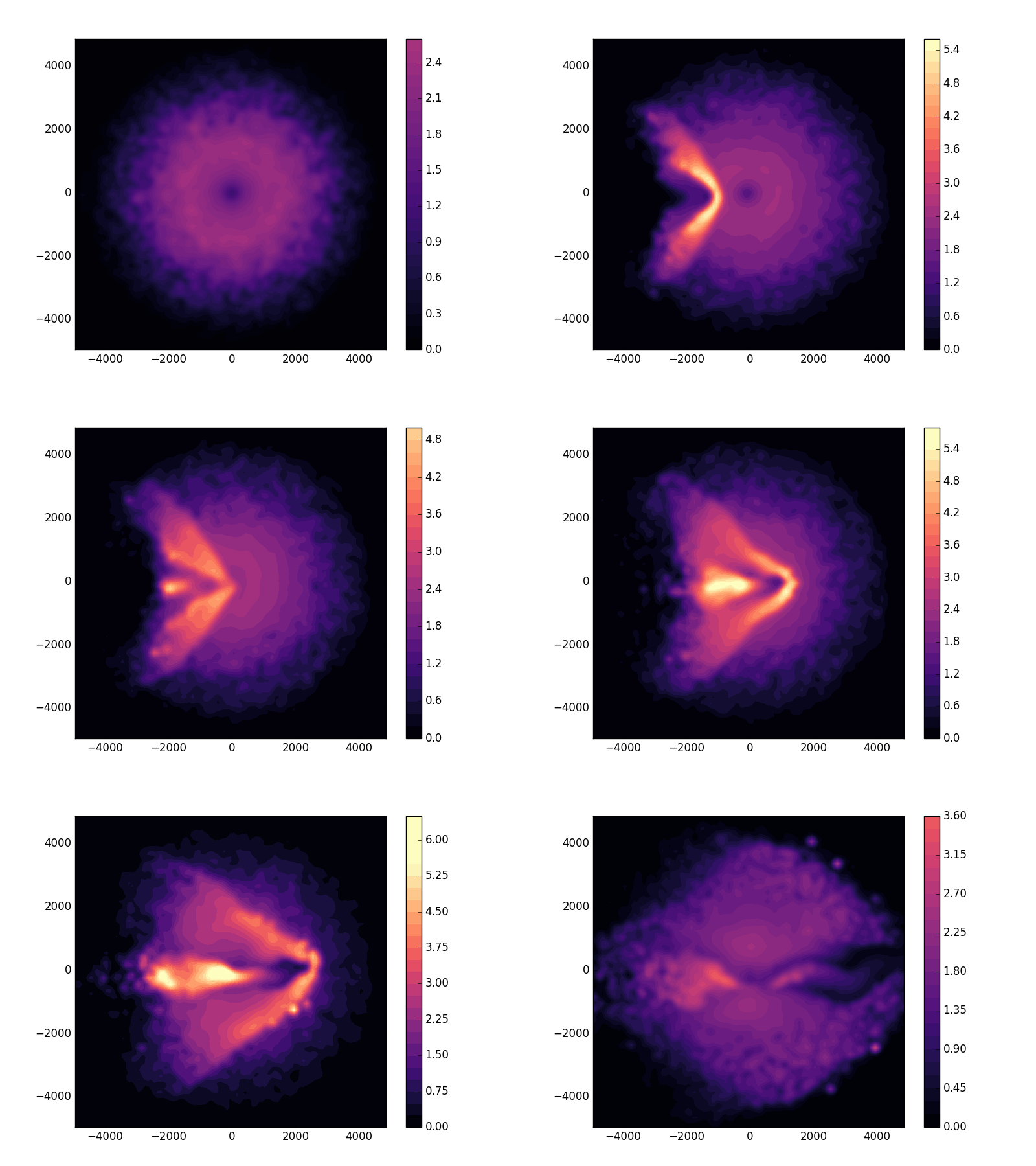}
   \caption{Internal energy of the dark plasma in a central slice of the Bullet Cluster collision. The epochs of the panels correspond to the same epochs in Fig.\ref{fig:videobullet}. All distances are in kpc, the internal energy per unit mass is given in units of 10$^6$ (km/s)$^2$. The energy scale is chosen such that the same colors correspond to equal temperatures in all panels.}
   \label{fig:bullet_T}
\end{figure*}

In Figs.~\ref{fig:abell_T} and \ref{fig:bullet_T} we plot the internal energy distributions of the dark plasma in the Abell 520 and the Bullet Cluster, respectively. The images show the slice of the collision plane.  The corresponding mass density distributions are depicted in Figs.~\ref{fig:videotrain} and \ref{fig:videobullet}. In the initial phase of both collisions, we observe a cone-shape feature with a hot interface and a cold interior. In later phases, in the Bullet Cluster the cone is seen to develop a hot center. This feature is not observed in the Abell 520 simulation.

\section{Conclusions}
\label{Conclusions}

In this article we simulated the mergers of the Abell 520 cluster and the Bullet Cluster under idealized conditions in the context of a two-component dark matter model, in which a fraction of dark matter is composed of particles with long range interactions mediated by a dark photon, while the rest remains non-interacting. The new DM component, dubbed dark plasma, is capable of forming shock-fronts that dissipate energy, thereby behaving similarly to the baryonic plasma in the intracluster medium during cluster mergers.

We performed a scan over the artificial viscosity parameter and the dark plasma fraction and found that the dark core observed in the Abell 520 cluster can be reproduced in the two-component model for a wide range of values for the artificial viscosity parameter $\eta$ if the fraction of dark plasma is at least 25-30 \%. The detailed control simulation shows that the interacting plasma component then forms Mach cones and a significant fraction remains in the center of the collision event. The Bullet cluster, on the other hand, does not exhibit such substructure. Indeed, we find that the dark plasma component of the smaller subcluster is partly absorbed by the larger halo during its core passage. The model thus remains compatible with the observation that the smaller subcluster has not lost more than 30\% of its mass during core passage. In conclusion, we find that the two-component model is sufficient to explain the observed features in both of these mergers.

Our results show that if dark matter is composed of two components with different collisional behavior, the commonly used measure of an offset between the center of mass and the luminosity peak of galaxies is not enough to capture the whole dynamics of self-interacting dark matter present in dissociative mergers. Instead, the dynamics of the merger event have to be carefully modeled case by case, as various outcomes are possible, depending on the initial dynamical configuration of the subclusters. To further test the two-component model more detailed simulations of the known cluster mergers are required, ideally also including the baryonic matter component in the simulation. Combined with increasing precision of the optical and X-ray observations of the merging clusters, it could be possible to identify features such as shock fronts in the dark matter substructure, which would serve as a smoking gun signature of the effective fluid nature of the interacting dark matter component.
Furthermore, the effects of the dark plasma during structure formation remain to be studied. In this work we have assumed the dark plasma component to initially follow an NFW halo profile with the structure supported by hydrostatic pressure. Cosmological simulations are required to determine if this is indeed the resulting structure from the collapse of primordial density perturbations, and to see what is the effect of the dark plasma for the abundance of substructure in the small scales.

\begin{acknowledgements}
The authors would like to thank Enn Saar and Gert H\"utsi for useful discussions. This research was supported by the Estonian Research Council via grants IUT26-2, IUT40-2, PUT799 and PUT716 and the ERDF CoE program via grant TK133. The work of M.H. has been supported by the Academy of Finland project number 267842. Figures \ref{fig:cluster_ratio}, \ref{fig:Clusters}, {\bf \ref{fig:abell_T} and \ref{fig:bullet_T}} were made using the {\sc matplotlib} \citep{Hunter:2007} python package.
\end{acknowledgements}

\bibliographystyle{aa}
\bibliography{mybib_aa_final}{}

\end{document}